# Ion-beam sculpting of nanowires

*Morteza Aramesh\**

Dr. M. Aramesh
Laboratory of Biosensors and Bioelectronics, Institute for Biomedical Engineering, ETH Zürich, Zürich 8092, Switzerland
Queensland University of Technology (QUT), Brisbane QLD 4000, Australia
E-mail: mrtz.aramesh@gmail.com; maramesh@ethz.ch

**Abstract:** Nanomaterials often undergo unusual mechanical deformations compared to their bulk counterparts when irradiated with ion-beams. This study visualizes and investigates some of the unusual interactions that can occur in nanomaterials during irradiation with medium-energy ion-beams using a Helium-Ion-Microscope (HIM). Ion-beam sculpting of semiconductor nanowires (NWs) with sub-10 nm features is demonstrated. Moreover, irradiation-induced growth of NWs at room-temperature is discovered. The new concept and possible mechanism of *irradiation-induced VLS* (vapor-liquid-solid) growth of NWs is introduced. These results are the basis for further fundamental and technological developments towards manipulation and visualization of ion-matter interactions at nanoscales.

Semiconductor nanomaterials hold outstanding potential as nano-components of future devices and systems such as lasers,[1] photodetectors,[2] single-electron memory devices [3] and sensors[4]. III–V nanowires (NWs) are particularly interesting due to their electrical and optical properties and also unique capability in quantum confinement.[5] III-V NWs, such as GaAs and InAs, are being widely implemented in optoelectronics research and industry.[6] III–V NWs can be fabricated on various substrates with potentially scalable and well-controlled methods such as metalorganic chemical-vapor-deposition (MOCVD) based on vapor–liquid–solid (VLS), [7, 8] vapor–solid–solid (VSS),[9] or via catalyst-free methods. [10] However, room-temperature growth of these NWs has not been demonstrated.

Controlled deformation and re-shaping of these nanowires may influence many of the state-of-the-art applications and open up new opportunities which were not possible to obtain before. [11] Previous studies have demonstrated that energetic ion-beams or electron-beams can be used as powerful tools to shape semiconductor nanowires.[12] Ion-beams with keV energy are particularly interesting for this purpose.[13-15] Recently, medium-energy light-ion irradiation of materials have gained attention, not only for capabilities for fabrication of well-defined nanostructure such as solid-state nanopores,[13, 16] but also for the ground they provide to study the ion-matter interaction at the nanoscales.[17]

Using a Helium-Ion-Microscope (HIM),[18] this study demonstrates that medium-energy ion-beam irradiation can be employed to *(i)* selectively and precisely shape the pre-grown III-V NWs and *(ii)* grow III-V NWs with ion-beam induced catalytic reactions at room-temperature. This study is divided into two parts. In *Part I*, it is demonstrated that irradiation of III-V NWs with $He^+$ can induce mass-flow via irradiation-induced diffusions. In *Part II*, the "irradiation-induced VLS" growth is presented, in which the concurring *irradiation-induced diffusion* and *irradiation-activated catalytic reactions* contribute into NW growth at room-temperature via VLS mechanism.

*Part I: Atomic-level diffusion and re-arrangement*

In the first part of this report irradiation-induced deformations in *free-standing* nanowires are studied in nano- and atomic-scales. HIM imaging allows direct observation of material response to the impinging ions at nanoscales. To demonstrate this capability, selective modification was realized by localized ion-beam irradiation at desired locations along the length of nanowires to achieve thinned areas with arbitrary diameters (**Figure 1.a,b** and **Figure S1**). Thinning can be done with potentially nanometer precision and interesting sub-10 nm structures can be obtained. High-resolution transmission electron microscopy (HRTEM) image shows that the crystal lattice of the NW was distorted at the irradiated region (**Figure 1.c** and **Figure S2**). NW thinning with ion-beam bombardment could be performed both for InAs and GaAs NWs at different $He^+$ ion-beam energies (10-45 keV).

Before the ion-beam bombardment the thickness of the nanowires was about 100 nm (see *e.g.* **Figure 1**), and later after the irradiation the thickness reduced significantly (< 10 nm). Based on SRIM simulations[19], the projected He$^+$ ion range for 30 keV ions in GaAs is much larger than the thickness of the nanowire. Therefore, most of the impinging ions (> 91%) are transmitted through the NW, and the ion-beam remains highly collimated inside the NW. With reduced thickness of the NW during the irradiation, the fraction of the transmitted ions increases significantly (*e.g.* > 99.7% for 10 nm-thick NW). This is accompanied with the very low efficiency of He$^+$ ions in sputtering of the atoms within the 30 keV range. The calculated sputtering yield is 0.14 and 0.11 atoms per ion, for 100 and 10 nm-thick NWs, respectively. Only 0.08-0.13% of the atoms are displaced during the irradiation with the corresponding ion-beam.

In the experiment which is shown **Figure S1**, an estimated volume of ~3.9×10$^5$ nm$^3$, (corresponding to ~1.4×10$^7$ atoms), was displaced from the NW after irradiation with fluence of ~ 10 ions.nm$^{-2}$. The amount of sputtered atoms (considering the sputtering is maximal) only corresponds to < 10% of volume reduction, which is not sufficient to explain the NW thinning. Therefore, the thinning of the NWs is not due to the ion-sputtering alone.

Indeed, as the irradiated area shrank to smaller diameter, the neighboring non-irradiated area became slightly thicker in diameter (**Figure S2**). A sort of mass-flow is therefore anticipated which allows transport of matter away from the irradiated region in NWs. The mechanism of mass-flow can be explained within the light of "ion-induced surface diffusion".[20] Generally, irradiation of low-dimensional materials generates axial stresses in the material[21], in which the relaxation is accompanied by atomic diffusion along the surface.[20] The "wind force"[21] of the ions is enough for generation of mass-flow. (Injection of "defects" – *e.g.* interstitials and vacancies by ion-beam can also promote surface diffusion and lateral mass-flow.[22]) Similar explanatory model of surface diffusion can be exploited to explain NWs thinning with He$^+$ ions. Upon irradiation with He$^+$, axial stress generates along the NW which promotes diffusion of the atoms away from the irradiation region and causes re-distribution of the matter. The existence of the axial stress within the NW is evident in the broken NWs, in which extensive deformations occurred immediately after the breaking moment (**Figure S1**).

It is concluded in this part, that medium-energy-ion-beams can be implemented to fashion III-V NWs at nano-dimensions via atomic re-arrangements, in which the mechanism could be understood by irradiation-induced diffusion model: the atoms of the irradiated materials are largely mobile and they are able to travel long distances along the surface.

*Part II: Irradiation-induced catalytic growth*

In this part ion-beam effects in *substrate-supported* nanowires are studied and also the new concept and possible mechanism of irradiation-induced VLS growth of NWs is introduced. Experiments on pre-grown NWs on substrates yielded the expected thinning with local ion-irradiation similar to free-standing nanowires (**Figure S3**). Surprisingly, I observed that the length and orientation of the substrate-supported pre-grown nanowires changed after irradiation (**Figure S4**). Remarkably, the wires grew in length and undertook dynamic orientations at different stages of their growth.

To understand the mechanism of NW growth with irradiation, the growth method was further extended to fabrication of NWs on samples with no pre-grown NWs, *i.e.* III-V substrate on which the Au particles were deposited and annealed prior to irradiation.[23] Interestingly, upon irradiation with He$^+$ ion beam, NWs started to grow at the locations of the gold particles. **Figure 2.a** shows the top surface of the sample before and after irradiation. Although the nucleation point of the NWs was defined by the location of the catalyst droplets, there were variations in the growth directions.

**Figure 2.b** shows early stages of NW growth in two different cases in which the irradiation conditions were the same. At the starting point of the growth, where the base of the NW column forms, minute differences can be detected in the location of the catalytic droplets relative to the bases of the NWs. But in the later stages, one can see that the two presented cases obviously look different from each other both in crystal facets and orientation angles. **Movie 1** shows an example for "randomness" in NW growth. Although nucleation randomness resulted in variations of growth process in later stages, the basal layer of the grown nanowires were mostly directed in (111) direction, maintaining epitaxial relationship with the substrate.

It was observed that by increasing the ion-beam energy and flux it is possible to achieve single-crystal NWs. **Figure 2.c** shows that NWs that were grown at 35 keV with higher currents (9-12 pA) exhibit single crystal growth characteristics. However, the growth rates were significantly slower than the previous attempts with lower fluxes and beam energies. A dose of ~$10^{17}$ ions.cm$^{-2}$ was required to grow a 1 µm long single-crystal NW. As observed in **Figure 2.c**, the high amount of the implanted dose can cause subsurface damages to the surface and substrate swelling,[14] which may not be ideal for some applications. Further refinements are still necessary to improve the precision, consistency and yield of the introduced NW growth process, but the main question to answer would be "What determines the evolution of NW during the irradiation induced-growth?"

He$^+$ is an inert gas and has extremely low reactivity with other materials. Like other noble gas atoms, He (or He$^+$) does not form bonds with atoms of other substances and as a result it quickly diffuses out from the implanted region, therefore no chemical reaction with He$^+$ ions is expected during the growth. Various aspects of irradiation-induced growth are analogous to the conventional VLS growth, except that: (*i*) the temperature is now only room-temperature, (*ii*) there is no external precursor gases (*iii*) the material is in its ionized state (due to irradiation[24]). This type of growth can be coined as "irradiation-induced VLS".

In a conventional VLS process, NWs are grown on a substrate as a result of gas phase interactions with a liquid catalyst. The process requires high-temperature annealing to allow wetting of the substrate with the molten catalyst (*e.g.* gold), which enables chemical interaction with precursor gases at the catalyst/substrate interface. [8, 25]

Two concurring effects contribute in room-temperature catalytic growth of NWs during bombardment with medium-energy ion-beams: *Irradiation-induced diffusion* of the substrate atoms and *Irradiation-induced viscosity* of gold/alloy droplet (**Figure 2.d**). The growth species are anticipated to be the migrating atoms from the substrate which are produced by irradiation-induced diffusion (as discussed in *Part I*). The mobility of the diffusive entities along the surface can be enhanced due to irradiation effects, and after the arrival of the diffusive species to the interface with the catalyst, bonding and attachment can occur at the available nucleation sites. Activating the catalytic reactions at the at Au/GaAs (or Au/InAs) interface is enabled by reducing the interface energy via reduced-viscosity. The catalyst droplet seems to be irradiation resistant to some extends, however due to extensive ionization the viscosity of the droplet is reduced and it can act like a liquid (although the metal droplet under irradiation may not completely transform to liquid-state).[22] The "liquid-like" droplet acts as a catalyst at the interface with underlying GaAs (or InAs) substrate and contributes in local growth of the NWs.

The size of the metal catalyst limits the lateral growth of the wire. Since the atoms of the metal catalyst are also gradually removed due to ion sputtering, the size of the catalyst reduces with time and therefore the resulted structure has a pyramid-like geometry (**Figure 2.a,b**). The variations in the orientation of nanowires could be influenced by ion-matter interactions at the nanoscales, among which instability of the ionized catalyst might be a possible cause for the fluctuations during growth.[26] Increasing the ion-beam energy and flux can potentially promote the charge-stability at the irradiated interfaces.[27] But since irradiation-induced damages on the substrate scales with ion-beam flux, it inversely influences the growth quality. Further challenging studies are required to understand and eventually control these unstable circumstances.

In conclusion, ion-matter interactions were visualized and investigated in a system which nanowires were irradiated with medium-energy He$^+$ ion-beams in a Helium-Ion-Microscope. It was demonstrated that growing and shaping of III-V nanowires at room temperature is possible by ion-beam irradiation. This is a simple and potentially scalable technique to grow III-V NWs at room temperature. A possible phenomenological mechanism was discussed to explain the observations. Ion-beam irradiation contributes in NW growth via reducing the interface energy at the catalyst/substrate interface and also by enhancing surface diffusion in the irradiated areas. Further studies with other materials and energy sources are required to fully realize the potentials of the presented technique for precise manipulation of NWs.

**Supporting Information**
Supporting Information is available from the Wiley Online Library or from the author.

**Acknowledgements**
I acknowledge Mun Teng Soo at University of Queensland for generous access to nanowire resources and TEM analysis. It is a pleasure to acknowledge discussions with Kostya Ostrikov, Josh Lipton-Duffin, Peter Hines, Annalena Wolff, Mun Teng Soo, Ivan Shorubalko, Tomaso Zambelli and János Vörös. I acknowledge CARF (Central Analytical Research Facility at Queensland University of Technology) Swiss Federal Laboratories for Materials Science and Technology (Empa) and Marie Skłodowska-Curie actions (Project Reference: 706930).

Received: ((will be filled in by the editorial staff))
Revised: ((will be filled in by the editorial staff))
Published online: ((will be filled in by the editorial staff))

**Figure 1.** Nanowire thinning due to irradiation with (30 keV) He⁺. (a) A (GaAs) NW was thinned at different locations on its length. Due to the high-resolution of HIM, the wires can be thinned to sub-10 nm dimensions (b). (c) HRTEM image of the thinned area, showing modifications in the crystal structure such as formation of crystal grains, voids and amorphous area.

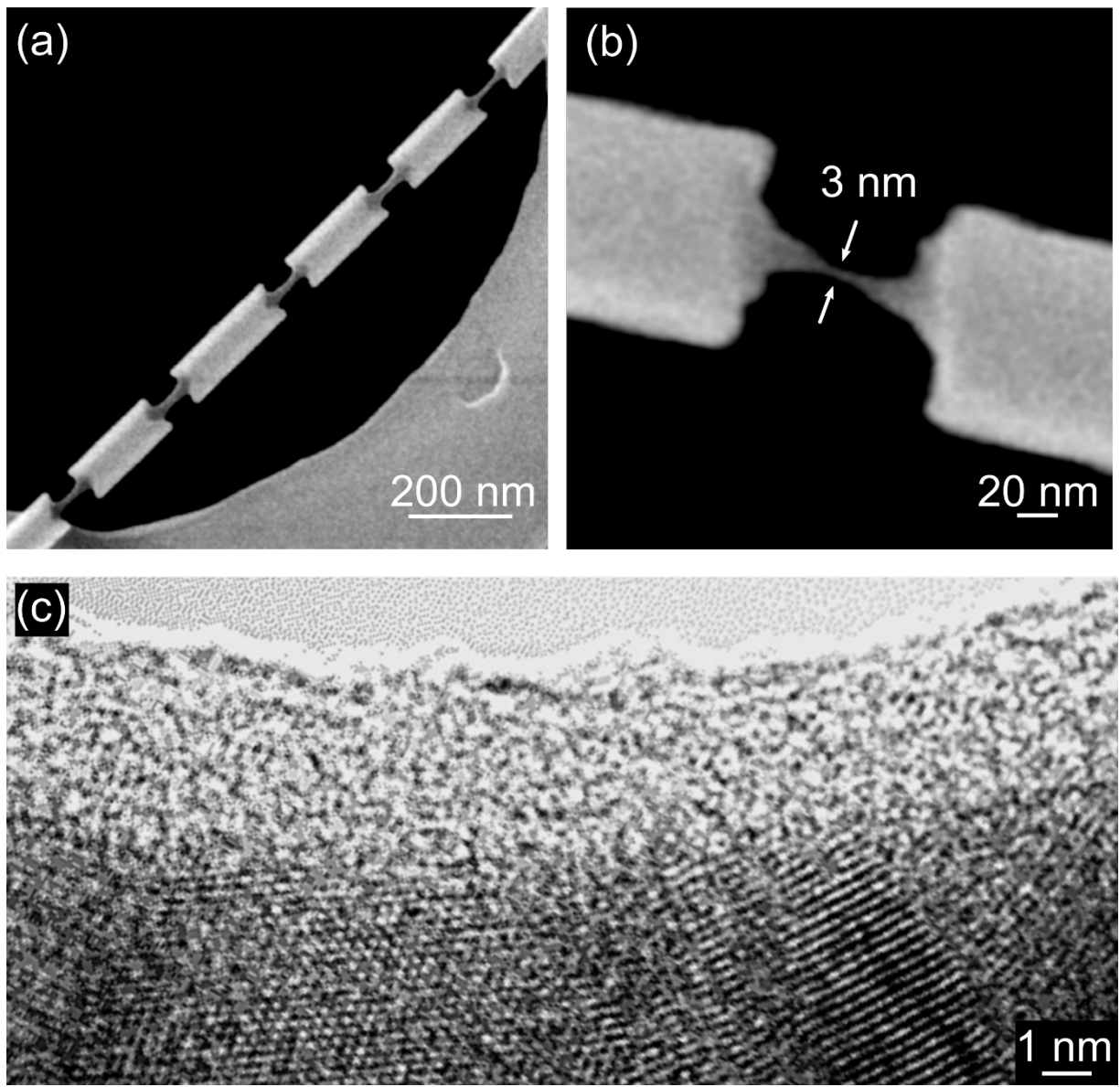

**Figure 2.** Room-temperature nanowire growth with medium-energy helium ion beams. (a) Schematic and micrographs of the sample before and after irradiation with He⁺. The Au composite acted as nucleation site for the growth of the NW induced by irradiation. (b) Shows the early stages of NW growth and randomness of the growth during the irradiation experiment. Initially, the two formed bases looked almost identical, but later each NW took a different orientation. (c) It is possible to grow single-crystal nanowires using this method. (d) shows schematic of the proposed mechanism for irradiation-induced VLS growth.

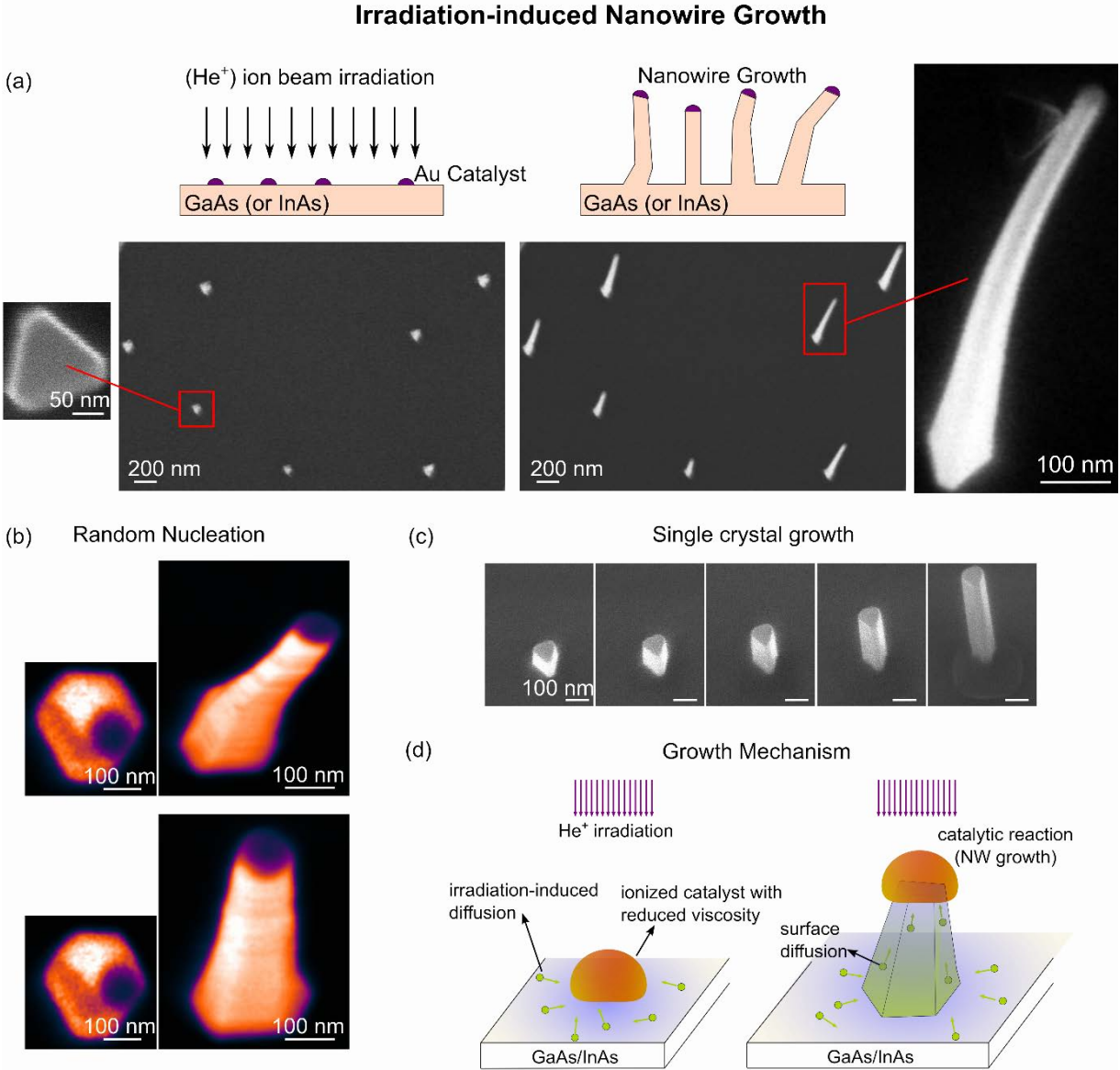

**Room-temperature growth and shaping of III-V nanowires** is essential for diverse applications. The recent progress made during the last decade towards nanoscale manipulation of nanomaterials with medium-energy ion-beams, has opened up new opportunities for fabrication of novel devices. Here, *Irradiation-induced VLS* is discovered which allows catalytic growth and shaping of III-V nanowires at room-temperature using helium ion-beams.

**Ion-beam sculpting of nanowires**

M. Aramesh*

**Irradiation-induced VLS:** Room-temperature growth and shaping of semiconductor nanowires by medium-energy ion-beam irradiation in a Helium-Ion-Microscope

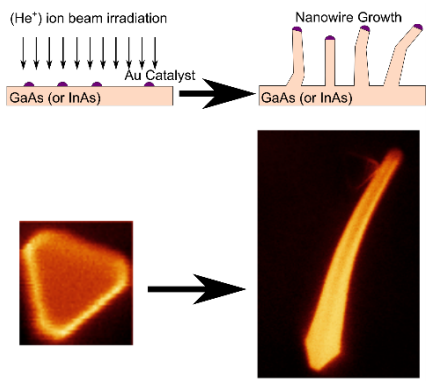